\documentclass[preprint,onecolumn]{elsarticle}
\usepackage{amssymb}
\usepackage{latexsym}
\usepackage[english]{babel}
\usepackage{times,mathptmx}
\usepackage{graphicx}

\begin{document}

\begin{frontmatter}

\title{Discussion of "Design of Controllers with Arbitrary Convergence Time" (Automatica 112 (2020) 108710)}

\author{Michael Basin}
\address{School of Physical and Mathematical
Sciences, Autonomous University of Nuevo Leon, San Nicolas de los
Garza, Nuevo Leon, Mexico.} \ead{mbasin@fcfm.uanl.mx}

%\author{Pablo Rodriguez-Ramirez}
%\address{School of Physical and Mathematical
%Sciences, Autonomous University of Nuevo Leon, San Nicolas de los
%Garza, Nuevo Leon, Mexico.} \ead{pablo.rodriguezrm@uanl.edu.mx}

%\fntext[label2]{The authors thank the Mexican National Science and
%Technology Council (CONACyT) for financial support under Grant
%250611.}

\end{frontmatter}

\textbf{Abstract.} This note corrects some technical inaccuracies in a recently published paper on predefined-time convergence \cite{Predef2020} and discusses implementation issues of the presented control algorithm.

\section{Introduction}

A recently published paper \cite{Predef2020} presents a novel predefined-time convergent control algorithm, where some technical inaccuracies are detected. The paper also claims that the designed control law is bounded even if the system state grows large. This note corrects the detected technical inaccuracies, so that the subsequent results of \cite{Predef2020} remain valid, and shows that a predefined-time convergent control law cannot be uniformly bounded for all state values. This note also discusses some implementation issues of the presented control algorithm, in particular, selection of gains in the backstepping procedure and location of the maximum values of the designed control inputs.

\section{Discussion}

\textbf{1.} In a recently published paper on predefined-time convergence \cite{Predef2020}, the non-autonomous differential equation
$$
\dot{x}=\frac{-\eta (e^x-1)}{e^x(t_f-t)},  \eqno (1)
$$
is considered with an initial condition $x(t_0)=x_0$, $t_0<t_f$, and $\eta >1$, whose solution is given by
$$
x(t)= \ln{(C(t_f-t)^{\eta }+1)},\ \mbox{ for }\ t\in [t_0,t_f], \eqno (2)
$$
$$
x(t)=0,\ \mbox{ for }\ t\ge t_f,
$$
where the integration constant $C = \frac{e^{x_0}-1}{(t_f-t_0)^{\eta }}$. It is further claimed (\cite{Predef2020}, Example in Introduction, 2nd paragraph): "It can be observed that
the right-hand side of (1) remains bounded even if $x$ grows large. This special feature makes the
system (1) perfectly suitable to function as a controller."

The last assertion is incorrect. Indeed, if $x<0$ and $\mid x\mid$ grows large, the term $e^x$ in the denominator of the right-hand side of (1)
tends to zero and the numerator tends to $\eta >1$, which has a finite nonzero value. Therefore, the whole right-hand side of (1) diverges to infinity at exponential rate.
This implies that the control law defined by the right-hand side of (1)
$$
u(t)=\frac{-\eta (e^x-1)}{e^x(t_f-t)} \eqno (3)
$$
becomes unacceptably larger at the initial time moment $t_0$ for large negative initial values $x_0<0$.

Fortunately, this situation can be corrected. Consider, instead of (1), the equation
$$
\dot{x}=\frac{-\eta (e^{\mid x\mid} -1)}{e^{\mid x\mid}(t_f-t)}sign{(x)},\quad  x(t_0)=x_0, \eqno (4)
$$
whose solution is given by
$$
x(t)= \ln{(C_1(t_f-t)^{\eta }+1)}sign{(x_0)},\ \mbox{ for }\ t\in [t_0,t_f], \eqno (5)
$$
$$
x(t)=0,\ \mbox{ for }\ t\ge t_f, 
$$
where $C_1 = \frac{e^{\mid x_0\mid }-1}{(t_f-t_0)^{\eta }}$. In this case, the absolute value of the right-hand side of (4) is bounded by $\frac{\eta}{t_f-t_0}$ for any $x_0$ at the initial time moment $t_0$, since $\frac{e^{\mid x\mid} -1}{e^{\mid x\mid}}$ tends to 1 as $x$ tends to infinity on any side, positive or negative. Note that the right-hand side of (4) is also continuous. After making this and similar corrections in subsequent parts of \cite{Predef2020}, the results of \cite{Predef2020} remain valid.

\textbf{2.} Note, however, that even the right-hand side of (4) does not remain bounded as $x$ grows large. Indeed,
the right-hand side of the equation (4) represents the velocity of convergence of the trajectory $x(t)$ from an initial
condition $x_0$ to zero. Therefore, if the convergence time is fixed as $t_f-t_0$, the velocity absolute value must be greater or equal than $\frac{\mid x_0\mid}{t_f-t_0}$
for at least one time moment in the interval $[t_0,t_f]$. Thus, the maximum absolute value of the right-hand side of (4) (and also (1)) diverges to infinity as the
absolute value of the initial condition $x_0$ grows. On the other hand, this is an expected price for predefined-time convergence, which must be independent of initial conditions.

\textbf{3.} For the purpose of practical implementation of the backstepping procedure proposed in \cite{Predef2020}, it is important to note that the gains $\eta_i$
should be selected greater than the system dimension, i.e., $\eta_i>2$ for $n=2$, $\eta_i>3$ for $n=3$, to avoid singularity (divergence to infinity) or discontinuity of the
backstepping control law at $t=t_f$. The singularity or discontinuity may appear in view of taking $(n-1)$ derivatives in time of the right-hand side of the equation (4) (or  (1)).

\textbf{4.} Finally, note the following feature of the control law defined by the right-hand side of (4) (or (1)). Recall the fixed-time convergent system
$$
\dot{x}= - k_1 \mid x\mid^{\alpha}sign{(x)} - k_2 \mid x\mid^{\beta}sign{(x)}, \quad  x(t_0)=x_0, \eqno (6)
$$
where $k_1,k_2>0$, $0<\alpha <1$, $\beta >1$. In this case, the absolute value of $x(t)$ decreases as time goes on from $t_0$; therefore, the maximum absolute value of the
control law defined by the right-hand side of (6) is registered at the initial time moment $t=t_0$. However, the maximum value of the right-hand side of (4) (or (1)) would be
reached at some intermediate time moment in the interval $[t_0,t_f]$, which can be checked by taking time derivative of the right-hand side of (4) (or (1)) and equating it to zero. This feature can be observed in Figs. 2, 3b, 4b, and 5b of \cite{Predef2020}, where the control inputs have their maximum values (which look like peaks in most cases) at intermediate time moments. Mathematically, this is a consequence of the explicit dependence on time of the non-autonomous right-hand side of (4) (or (1)). From the viewpoint of practical implementation, lower initial values of the control law may be confusing for correct evaluation of the required control magnitude.

\bibliography{2018ANR}{}
\bibliographystyle{elsarticle-num}

\end{document}